\documentstyle[11pt,aaspp4]{article}

\newcommand\etal{et al. }
\newcommand\eg{{e. g}. }
\newcommand\ie{{\it i. e}. }
\newcommand\be {\begin{equation}}
\newcommand\en{\end{equation}}

\slugcomment{Submitted to {\it The Astrophysical Journal}}

\begin{document}

\title{THE PARKER INSTABILITY IN A THICK GASEOUS DISK II: NUMERICAL
       SIMULATIONS IN 2D}     

\author{Alfredo Santill\'an\altaffilmark{1,2}, Jongsoo Kim\altaffilmark{3},
        Jos\'e Franco\altaffilmark{2}, Marco Martos\altaffilmark{2},
        Seung Soo Hong\altaffilmark{4} \and Dongsu Ryu\altaffilmark{5}}

\altaffiltext{1}{C\'omputo Aplicado--DGSCA, Zona Cultural, Universidad Nacional 
Aut\'onoma de M\'exico, 04510 M\'exico D.F., M\'exico; Electronic mail: 
alfredo@astroscu.unam.mx}
\altaffiltext{2}{Instituto de Astronom\'\i a, Universidad Nacional Aut\'onoma 
de M\'exico, A. P. 70-264, 04510 M\'exico D.F., M\'exico; Electronic mail: 
pepe@astroscu.unam.mx, marco@astroscu.unam.mx}
\altaffiltext{3}{Korea Astronomy Observatory, 61-1, Hwaam-Dong,
Yusong-Ku, Taejon 305-348, Korea; Electronic mail: jskim@hanul.issa.re.kr}
\altaffiltext{4}{Department of Astronomy, Seoul National University, Seoul
151-742, Korea; Electronic mail: sshong@astroism.snu.ac.kr}
\altaffiltext{5}{Department of Astronomy \& Space Science, Chungnam National 
University, Daejeon 305-764, Korea; Electronic mail: ryu@canopus.chungnam.ac.kr} 
                           
\begin{abstract}  
We present 2D, ideal--MHD numerical simulations of the Parker instability in a 
multi--component warm disk model. The calculations were done using two numerical codes
with different algorithms, TVD and ZEUS-3D. The outcome of the numerical 
experiments performed with both codes is very similar, and confirms the results 
of the linear analysis for the undular mode derived by Kim \etal (2000): the most 
unstable wavelength is about 3 kpc and its growth timescale is between 30--50 Myr 
(the growth rate is sensitive to the position of the upper boundary of the numerical 
grid). Thus, the time and length scales of this multicomponent disk model are 
substantially larger than those derived for thin disk models. We use three different 
types of perturbations, random, symmetric, and antisymmetric, to trigger the instability. 
The antisymmetric mode is dominant, and determines the minimum time for the onset of 
the nonlinear regime. The instability generates dense condensations and the final peak
column density value in the antisymmetric case, as also derived by Kim \etal (2000), is
about a factor of 3 larger than its initial value. These wavelengths
and density enhancement factors indicate that the instability alone
cannot be the main formation mechanism of giant molecular clouds in
the general interstellar medium. The role of the instability in the
formation of large-scale corrugations along spiral arms is briefly discussed.
 
\end{abstract}
  
\keywords{instabilities --- ISM: clouds --- ISM:
magnetic fields --- ISM: structure --- MHD}

\section{INTRODUCTION}

The interstellar disk in our Galaxy is thicker and more magnetized than previously
assumed, and the transition between the gaseous disk and the halo has a complex 
magnetic field structure and extended gas components, including the ionized Reynolds 
layer (Hoyle \& Ellis 1963; Reynolds 1989). This type of extended disk also seems to be  
common in other spiral galaxies, and thick diffuse gas layers, with worms and dusty 
filaments, have been observed, for instance, in NGC 891, NGC 5775, and NGC 4302 (Dettmar
1990; Rand, Kulkarni \& Hester 1990; Rand 1995; Howk \& Savage 1997). Similarly, the
observed magnetic fields in edge--on galaxies can be traced well above the main gaseous
disk (Hummel \& Beck 1995). 

Boulares \& Cox (1990; hereafter BC) have incorporated the observed extended gas layers,
the gradient of the magnetic field, and the cosmic ray pressure in a model for the
thick disk of the Galaxy (see also Kalberla \& Kerp 1998). The existence of magnetized gas
layers with large scale heights has important consequences for the structure of the gaseous
disk, and its response to large scale perturbations (see Cox 1990; Martos 1993; Franco
\etal 1995). For instance, the total interstellar pressure in a disk with extended layers
is substantially larger than that for thin disk models. Also, the interaction
between the disk and the halo is more complex: spiral density waves can produce flows,
shocks, and even star formation at high galactic latitudes (Martos \& Cox 1998; Martos
\etal 1999), whereas the penetration of high--velocity clouds into the disk is severely
reduced by magnetic pressure and tension (\eg Santill\'an \etal 1999). The stability of
the system is also modified and, depending on the temperature and ${\bf B}$-field
distributions, the equilibrium configuration of a thick disk can be Parker unstable. For
the observed conditions at the solar neighborhood (BC), the linear stability analysis of a
multi-component isothermal disk indicates that the fastest growing undular mode has a
wavelength of the order of 3 kpc (Kim \etal 2000; hereafter Paper I). This is a factor of
8 larger than the one derived for the thin gaseous disk, and illustrates the variations
that can be expected when considering the extended gas layers.

In Paper I we derived both the dispersion relations and the final equilibrium configuration
in two-dimensions (2D)
of the undular mode of the Parker instability. This mode generates high-density
condensations at the magnetic valleys, which are eventually stabilized by magnetic tension
and the system settles down into a new final equilibrium. The linear analysis provides the
initial growth rates and inter--distance scales of the condensations, and the final
equilibrium configuration (obtained by solving the magneto-hydrostatic equations with
the flux freezing condition) provides their final steady-state structure. To complete the
study, here we present 2D numerical simulations for the evolution of the instability. This
enables us to verify the results of the linear analysis, and to follow the details of the
non-linear regime until the final equilibrium is reached.  
As a check on the reliability of
the numerical results, the numerical experiments have been performed with two different MHD
numerical codes, ZEUS-3D and TVD. The results obtained with both codes are very similar,
and confirm the growth rates and wavelengths of the fastest growing modes derived in Paper
I. The plan of the present paper is as follows. The details of the numerical work are
described in the next section. The results are described in Section 3, and a brief summary
and discussion are given in Section 4.

\section{NUMERICAL SETUP AND CODES} 

Since the first one-dimensional simulation of the problem (Baierlein 1983), there has 
been a number of numerical studies of the evolution of the Parker instability under
different conditions. For instance, Matsumoto \etal (1988, 1990) and Matsumoto \& Shibata
(1992) have explored the nature of the undular and mixed modes for accretion disk
environments with 2D and 3D simulations. Also, 
using the same numerical code, Shibata \etal (1989a, 1989b), Kaisig \etal (1990) and
Nozawa \etal (1992) have used the Parker instability as the driving mechanism for 
emergent flux tubes in the Sun, and explained several features of the solar activity. In 
a different type of study, addressing the question of giant molecular cloud formation,
Basu \etal (1997) and Kim \etal (1998) have performed 2D and 3D simulations of the
instability in a thin gaseous disk under the influence of a simplified uniform gravity.   
Here we follow the evolution of the undular instability in a multi-component, isothermal 
and magnetized gaseous disk, with the properties of the solar neighborhood. 

\subsection{A Conservative Form of the Isothermal MHD Equations}

We use two-dimensional Cartesian coordinates $(y,~z)$, whose directions are azimuthal and 
perpendicular to the Galactic disk, respectively. When the vertical gravitational acceleration
has only $z$-component, ${\bf g} = -g(z)\hat{z}$, the conservative form of the isothermal 
MHD equations is
(see Paper I)
\begin{equation}\label{imhd}
\frac{\partial {\bf q}}{\partial t}
+ \frac{\partial {\bf F_y}}{\partial y}
+ \frac{\partial {\bf F_z}}{\partial z} = {\bf S},
\end{equation}
\begin{equation}\label{q}
{\bf q} = \left(\begin{array}{c}
                  \rho     \\
                  \rho v_y \\
                  \rho v_z \\
                  B_y      \\
                  B_z
                \end{array}
          \right),
\end{equation}
\begin{equation}\label{f}
{\bf F_y} = \left( \begin{array}{c}
                     \rho v_y                                 \\
                     \rho v_y^2 + a^2 \rho + \frac{1}{8\pi}(B_z^2-B_y^2)  \\
                     \rho v_y v_z - \frac{1}{4\pi} B_y B_z                \\
                     0                                        \\
                     v_y B_z - v_z B_y                        \\
                   \end{array}
            \right), \;
{\bf F_z} = \left( \begin{array}{c}
                     \rho v_z                                 \\
                     \rho v_y v_z - \frac{1}{4\pi} B_y B_z    \\
                     \rho v_z^2 + a^2 \rho + \frac{1}{8\pi}(B_y^2-B_z^2)  \\
                     v_z B_y - v_y B_z                        \\
                     0                                        \\
                   \end{array}
            \right),
\end{equation}
\begin{equation}\label{s}
{\bf S} = \left( \begin{array}{c}
                     0      \\
                     0      \\
                     -g v_z \\
                     0      \\
                     0      \\
                 \end{array}
          \right),
\end{equation}
where $a$ is the isothermal speed for all interstellar components (\ie the velocity 
dispersion of all gas components is set to the same value), and the rest of the 
symbols have their usual meaning.  

\subsection{Initial Condition: The Thick Galactic Disk}

As the initial condition for our numerical experiments, we introduce the Galactic 
disk model that was originally proposed by Martos (1993). The model is the  
magnetohydrostatic equilibrium of a system composed of five interstellar gas
layers, with a horizontal magnetic field and a realistic gravitational acceleration. 
Since the detailed building procedure of the model is given in Santill\'an et al.~(1999)
and Paper I, here we give only a brief description of the model.

The two building blocks of the model are the density distribution of the multi-component
gaseous disk (BC),
\begin{eqnarray}
\label{eq:n}
n(z) &=& 0.6 \exp \left[ - \frac{z^2}{2(70 \mbox{pc})^2} \right]
       + 0.3 \exp \left[ - \frac{z^2}{2(135\mbox{pc})^2} \right]
       + 0.07\exp \left[ - \frac{z^2}{2(135\mbox{pc})^2} \right] \nonumber \\
     &+& 0.1 \exp \left[ - \frac{|z|}{400\mbox{pc}} \right] + 0.03 \exp
       \left[ - \frac{|z|}{900\mbox{pc}}\right]~~~{\rm cm}^{-3},
\end{eqnarray}
and a simple fitting formula (Martos 1993) to the gravitational acceleration 
at the solar neighborhood (Bienaym\'e, Robin \& Cr\'ez\'e 1987),
\begin{equation}
\label{eq:gravity}
g(z) = 8 \times 10^{-9}
       \left[1 - 0.52 \exp\left(-\frac{|z|}{325 \mbox{pc}} \right)
               - 0.48 \exp\left(-\frac{|z|}{900 \mbox{pc}} \right)
       \right]~~~{\rm cm~s}^{-2}.
\end{equation}
The mass density stratification, including 10\% He by number, is $\rho(z) = 1.27 m_{\rm 
H} n(z)$, where $m_{\rm H}$ is the mass of a hydrogen atom. The total (gas plus magnetic)
pressure, $P_0(z)$, is obtained from the integral of $\rho(z)$ and $g(z)$ with the
boundary condition, $P_0 (z=10\ {\rm kpc})=0$. Assuming that the gas is isothermal and
setting the midplane value of the magnetic field, $B_0(0) \simeq 5\mu$G (BC; Heiles 1996),
we deduce the initial field stratification, $B(z)$, from the
distribution of the total pressure $P_0(z)$.   
In contrast to thin disk models, where the magnetic-to-thermal pressure ratio $\alpha$ is usually
held constant, our isothermal thick disk has a $z$-dependent $\alpha$ that increases with
height (see Paper I). Thus, the high-latitude gas is largely supported by the magnetic
field, and is unstable to the undular mode. 

The isothermal thick disk model is specified by two parameters: the effective scale 
height, $H_{\rm eff}=166$ pc, and the isothermal sound speed $a=8.4$ km s$^{-1}$ (see 
Paper I). This scale height is simply defined as half of the total disk column density
divided by the density at the midplane, and the sound speed is determined by the midplane
values $P_0(0)$ and $B_0(0)$. These two parameters, $H_{\rm eff}$ and $a$,
define our units of length 
and velocity. The time unit is defined as $H_{\rm eff}/a=1.9 \times 10^7$ yr and, unless
otherwise stated, all physical quantities are normalized to these three units.

\subsection{Perturbations}

The linear stability analysis in Paper I showed that the above equilibrium state 
is unstable against the undular instability when the perturbation wavelength is longer 
than about 1.5 to 1.8 kpc, depending on the location of the nodal point (\ie the location
where $v_z=0$; see below). These critical wavelength values correspond to nodal points
located between 5 and 1.5 kpc above midplane, respectively. Thus, in order to initiate the
instability in the numerical experiments, the equilibrium state is perturbed with
appropriate wavelengths. 

We have used three kinds of perturbations. The first type is the
even parity midplane symmetric (MS) perturbation, which forms structures distributed
symmetrically with respect to the midplane, 
\begin{equation}
\label{eq:MSpert}
v_z = C \sin(\frac{2 \pi y}{\lambda_y}) \sin(\frac{2 \pi z}{\lambda_z}), 
\end{equation}
where $C$ is the amplitude of the vertical velocity perturbation, and $\lambda_y$ and
$\lambda_z$ are the horizontal and vertical wavelengths of the perturbations. The second
type is the odd parity midplane antisymmetric (MA) perturbation, 
\begin{equation}
\label{eq:MApert}
v_z = C \sin(\frac{2 \pi y}{\lambda_y}) \cos(\frac{2 \pi z}{\lambda_z}).  
\end{equation}
The initial amplitudes for both MS and MA perturbations are usually set
to $C= 10^{-3}a$. 

The last type is a random velocity perturbation, without any preferred
wavelength or parity. The standard deviation of each velocity component
is here set to 10$^{-4}a$.  

\subsection{Boundary Conditions}

In deriving the dispersion relations in Paper I, perturbations were allowed to vary
periodically along the horizontal axis.
The boundary conditions at midplane determined
the parity of the perturbation (\ie $v_z=0$ at $z=0$ for the MS perturbation,
and $dv_z/dz=0$ at $z=0$ for the MA case).
However, regardless of parity, the vertical velocity perturbation at the upper
(or lower) boundary was always set equal to zero. Therefore, the upper boundary becomes
a nodal point.

In the numerical experiments, for consistency with the boundary conditions used in the
linear analysis, we use a periodic condition at the $y$-boundaries of the computational
domain, and a reflecting condition at the $z$-boundaries.  The reflecting condition
imposes $v_z=0$ on the vertical boundaries, which means that the
boundaries mimic the nodal points of the linear analysis. Given that no
flow is allowed through the $z$-boundaries and any outflow at one
horizontal boundary is always compensated by an identical inflow at
the other boundary, the mass in the computational domain is conserved
at all times. This allows us to easily track the evolution of the
energy content of the system.

The even parity perturbation is symmetric with respect to the midplane
and has a nodal point at $z=0$. Thus, $\lambda_z$ in equation
(\ref{eq:MSpert}) should be set equal 2$z_{\rm node}$, and it is
sufficient to use only one (the upper) hemisphere as the computational
domain. For the case of an odd parity perturbation, $\lambda_z$ in
equation~(\ref{eq:MApert}) should be set equal to
4$z_{\rm node}$, and the simulations have to be performed with both
hemispheres in the computational domain (see Table~1).

For completeness, we have also performed simulations with either outflow or
inflow conditions at the $z$-boundaries. The results for the fastest
growing modes are similar to ones presented here but, given that the
magnetic and velocity fields change when the gas is allowed through the
$z$-boundaries, some differences appear for perturbations with longer
wavelengths. These results will be presented in a future communication.

\subsection{Numerical Codes}

The evolution of the undular instability is followed with both the isothermal MHD-TVD 
code described by Kim et al. (1999), and ZEUS-3D which has been described by Stone \&
Norman (1992a, 1992b). The first code is based on a second-order accurate, conservative,
explicit, total variation diminishing (TVD) method. The $\nabla \cdot {\bf B}=0$ 
condition is maintained during the simulations by a scheme similar to a constrained
transport (CT) scheme (Evans \& Hawley 1988), which is reported in Ryu et al. (1998). The
ZEUS-3D code, on the other hand, solves the ideal MHD equations on an Eulerian staggered
mesh using an operator-split formalism wherein accelerations and heating are computed
using centered differences followed by a conservative step. Monotonic upwind interpolation
of either second or third order (van Leer or PPM advection) is used. Shocks are smeared
using artificial viscosity. The CT algorithm is also adopted for the evolution
of the magnetic field. 

The main differences between the two codes are: i) the TVD code uses an approximate Riemann
solver whereas ZEUS-3D uses a finite difference algorithm, ii) the present version of the
TVD code solves the isothermal MHD equations (without the energy term) whereas ZEUS-3D
has the adiabatic MHD with $\gamma=1.01$, and iii) ZEUS-3D allows less restrictive boundary
conditions. The use of two different numerical algorithms, together with the linear
analysis in Paper I, allows for a verification of the details of the numerical results. As
described in the next section, and except for secondary details, the results from both
codes are nearly identical. The advantage we have found in the isothermal TVD code is
that it uses less cpu time per run. This is an expected advantage, however, because the
code was optimized for isothermal problems and does not solve the energy
equation. The important result of this exercise, then, is
that both codes handle the problem in an adequate manner and with similar accuracies. 

\subsection{The Set of Numerical Experiments}

In Table~1 we summarize the parameters of the numerical experiments presented in this
paper. The simulations can be grouped into three categories.  The first group is a set of
$(1+\frac{1}{2})$D ``equilibrium tests'' (see below), performed with different resolutions
and labeled as cases 1, 2, and 3. They were designed to check whether the codes were able
to maintain the initial magnetohydrostatic equilibrium over long times. Within the accuracy
of the algorithm for solving the equality in eqn.~(\ref{imhd}), a small force unbalance is
allowed at every time step and its accumulative effect may affect the long-term evolution
of the instability. 

The second group of experiments is designed to compare the evolution, growth rates, and
final equilibrium states for perturbations with different parities. This group is labeled
as cases 4, 5, and 6. For this purpose, we have perturbed the initial state with MS and MA
perturbations that have selected values of $\lambda_y$ and $\lambda_z$. The cases shown
here have $\lambda_y$ equal to the most unstable wavelength and, as stated above,
$\lambda_z$ is equal to an integer number of times $z_{\rm node}$.
  
The third group is illustrated with only one simulation, case 7, and is devised to pick up
the value of the most unstable wavelength and the preferred parity (MS or MA) dominating
the system. For this purpose, we generate random velocity perturbations in an extended
computational domain. 
 
\section{RESULTS}

As stated before, the coordinates ($y$, $z$) represent distances along and perpendicular
to the midplane, respectively. The $y$-axis is anchored at the solar galactocentric radius
and is quasi-azimuthal, defined by the tangent of the local field orientation in our
Galaxy (see Heiles 1996 and Valle\'e 1997)

\subsection{Equilibrium Tests: Thick Disk Oscillations}

By $(1+\frac{1}{2})$D we mean that: i) the state vector ${\bf q}$ in eqn.~(\ref{imhd})
is a function of only $z$ and $t$, so $\partial {\bf F_y}/\partial y=0$, but ii) the
$y$-components of ${\bf v}$ and ${\bf B}$ survive and vary with $z$. In 
the isothermal TVD code, the source term of the gravitational acceleration is separated
from the rest of the MHD equations by use of a splitting procedure (a detailed description
of this procedure is given in Kim \etal 1999), but in ZEUS-3D  
the term is incorporated into the MHD equations (Stone \& Norman 1992a,b). 
Here we verify if these different methods are able to maintain the magnetohydrostatic
equilibrium over long periods of time. 

After setting the initial condition described in \S 2.2, we trace the maximum vertical
velocity as a function of time (Fig.~1). Numerical noise is produced by slight force
imbalances at every cell, with wavelengths comparable to the cell-size, and propagates along the
computational domain as compressive magnetosonic waves. The waves propagate through the
stratified medium, and reach their peak velocities near the $z$-boundaries. The response of
the gaseous disk to the continuous perturbation generated by this numerical noise is to
oscillate with the natural frequency of the system. Obviously, the velocities
involved in the normal mode of
oscillation vary with height, from the ''shallow water wave'' behavior near the plane to the
more complex response at higher values of $z$ (see Martos 1993 and Martos
 \&
Cox 1998). The harmonic motion period in the linear gravity regime (for
$z$ within 170 pc from midplane) is proportional to
\begin{equation}
\left( \frac{\partial g(z)}{\partial z}\right)^{-1/2}|_{z=0} \simeq 1.4 \times 10^7
\ {\rm yr},
\end{equation}
and such approximation breaks down at high latitudes. Thus, the system acts as a complex set of
coupled oscillators, and a modulated pattern should appear in our thick disk. This is
indeed seen in the equilibrium tests performed with both codes. 

In case~1, the simulation with the coarsest resolution, the peak velocity value shows
modulated oscillations with a very small amplitude during the whole evolution. The
resulting amplitude decreases to even lower values as the resolution is increased in cases
2 and 3. The top panel of Figure 1 shows the results for case 1 performed with ZEUS-3D. The oscillation
amplitude is always less than 1\% of the isothermal sound speed, and the plot shows the
modulated oscillatory pattern. The oscillation is complex but three  components
with periods of about $2.8\times 10^7$, $6\times 10^7$ and $11\times 10^7$ year are revealed. 

The lower panel of Figure 1 shows the same case performed with the
isothermal TVD code. Since the TVD code is based on the Riemann
solution, the gravity term should be treated separately. The code first
updates the fluid variables (density, magnetic field, and velocity)
without taking into account the gravitational acceleration. The force
unbalance generated by this first step is then compensated by adding
the vertical momentum due to gravity. There are residuals after this
second step, because the code calculates the difference between two 
large numbers. Thus, due to a transient generated by the initial
residuals, the first few velocity peaks are slightly larger than those
obtained with ZEUS-3D, but the oscillation amplitude decreases
below 1 \% of the sound speed for the rest of the evolution. Also, the
plot shows a cleaner oscillation mode, with a period of about $6\times
10^7$ year, because a fixed velocity pattern was enforced at the
inner parts of the disk (\ie this acts like a filter; the central parts
of the disk are not allowed to oscillate at their natural frequencies,
and one can isolate the oscillations of the extended layers). 

The main conclusion of these tests, then, is that both ZEUS and TVD
preserve the equilibrium state within an accuracy of a maximum spurious
velocity of less than 1\% after an elapsed time of about $10^9$ yr.

\subsection{Sinusoidal MS and MA Perturbations} 

As stated above, cases 4, 5 and 6 have been constructed to follow the evolution of the
undular instability triggered by MS and MA perturbations. From Paper I, the wavelength with the
maximum growth rate in our thick disk model is about 18$H_{\rm eff}$, and this value is
not sensitive to the parity of the perturbation. The three cases shown here have been
done with this wavelength value but varying, aside from the perturbation
parity, the domain size and grid resolution (see Table 1).  

The four panels of Figure~2 show the rms value of the horizontal (dotted lines) and
vertical (dashed lines) velocities as a function of time for both MS (left panels) and 
MA (right panels) perturbations. The solid line has a slope equal to the growth rate for
the fastest growing mode with $z_{\rm node}= 9$ derived in Paper I. The results
obtained with the TVD code are displayed in the two upper panels, and the ones from ZEUS-3D
are given in the lower panels. As mentioned before, the rms velocities are given in units
of the isothermal sound speed (the panels have a natural logarithmic scale), and the units
of time are $H_{\rm eff}/a$. The plots show a good consistency in the results attained with
both codes. 

The shapes of the velocity curves suggest that the evolution of the undular instability
can be divided into three well defined stages: a linear growth, a nonlinear saturated
regime, and a damping oscillatory stage. During the linear part (up to $t=35$ for the MS
mode, and up to $t=27$ for the MA mode), the rms velocities increase at a constant growth
rate. At early times, before $t=20$ for the MS mode and $t=15$ for the MA mode, the
vertical rms velocity simply oscillates. The horizontal rms velocity, however, grows nearly
steadily from zero (after a few oscillations) to its maximum value. A close look at the
slopes for the horizontal rms velocities indicates that the slope of the MA case is
slightly steeper than that of the MS case. This is a manifestation that the MA mode is
slightly more unstable than the MS mode (Horiuchi \etal 1988). We did not derive this in
the linear analysis of Paper I, but here show numerically that it is indeed the case in the
next subsection. The end time for this phase can be defined as the
moment when, due to the action of magnetic tension, the growth rate decreases abruptly to
zero ($t=35$ for MS, and $t=27$ for MA). In a more strict sense, the term ``linear'' should
be applied to times when the velocity has only a modest increase, smaller than the sound
speed. Nonetheless, the rms values for the horizontal velocity at the end of the linear
stage are nearly equal to the sound speed. This implies that the ``linear approximation''
in this case goes well beyond its conventional limits.

The end of the linear phase marks the beginning of the nonlinear saturated stage, in which
the rms velocities reach their peak values. These peak values are maintained for a brief
period of time, and then the velocities drop down and begin to oscillate about a nearly
constant final value. It is a matter of taste to define the end epoch of this nonlinear
stage. For simplicity, we take this moment as the time when the first oscillation begins,
$t=55$ for the MS case and $t=45$ for the MA case. This corresponds to, approximately,
the moment when the gas accumulated in the ``magnetic valleys'', which is compressed by the
ram pressure of the gas falling along the distorted field lines, begins to re-expand. After
a large fraction of the halo gas has fallen into the valleys, the ram pressure decreases
rapidly and the condensations readjust. The system also enters into its final
equilibrium stage: as the rate of mass falling into the valley stalls, the weight of the
condensations is stabilized and the tension of the distorted field lines is able to
support them. Thus, the gas structures begin to stabilize in the damping oscillatory
phase. As time goes by, the rms velocity gradually decreases due to numerical dissipation,
and this is the reason why we use the term ``damping'' (see Kim \etal 1999).

Figure 3 shows snapshots of the density structures, velocity fields, and magnetic field
topologies at three selected times. The snapshots obtained with both
codes are nearly identical and, for simplicity, we show only the ones
obtained with the TVD simulations.
The left panels show case~4 with the MS perturbation,
and the right panels show case~5 with the MA mode. The velocity fields are represented
with arrows, and the standard velocity vector (with size $2a$) is shown at the top of the
left panels. The $B$-field lines are chosen in such a way that the flux between
consecutive lines is constant. The densities are color-coded from red to purple, as the
value decreases. The initial sinusoidal perturbations propagate vertically in the form of
magnetosonic waves and, as stated above, reach their peak values near the $z$-boundaries of
the computational domain (see also Fig.~1).

The two upper panels, corresponding to the end time of the linear stage ($t=35$ for the MS
case, and $t=27$ for the MA case), show that the instability is already fully developed.
The midplane is slightly bended in the MA case. Later on, during the nonlinear stage, gas
with supersonic velocities is continuously accumulated into the magnetic valleys (some
weak shocks are generated during this stage), forming
spur-like structures (see middle panels). The ram pressure of the falling gas keeps the
layer thin and compressed, creating relatively large central densities. At the end of the
nonlinear stage, the ram pressure drops and the over-compressed structures become thicker.
The gas velocity near the edge of the spurs reverses its direction, and the system enters
into the damping oscillatory state. The structures eventually move into a quasistatic
state (lower panels), which is similar to the final equilibrium state obtained in Paper I.
 
The runs of the different energies in the system are plotted in Figure~4. The left side
corresponds to the MS case, and the right side to the MA case. The internal, kinetic,
magnetic, gravitational and heat-exchange energies are defined by 
\begin{equation}
E_i = \int\!\!\int \frac{3}{2}\;p\;dy dz, \ \ \
E_k = \int\!\!\int \frac{1}{2} \rho (v_y^2+v_z^2) dydz,
\end{equation}
\begin{equation}
E_m = \int\!\!\int \frac{1}{8\pi} (B_y^2+B_z^2) dydz, \\\ 
E_g = \int\!\!\int \rho\;\phi\; dydz, \ \ \ 
E_h = \int\!\!\int\;p\;\ln p\; dydz,  
\end{equation}
where $\phi=\int_0^z g(z)$ is the gravitational potential, and the double integrals are
performed over the whole computational domain. Due to the isothermal condition, the internal
energy is just proportional to the total system mass. Since there is no mass exchange
between our system and the outside, the total mass is conserved in our simulations (see
solid lines, labeled $E_i$). We use the internal energy as the normalization unit for all
energies.

Following Mouschovias (1974), we define the energy integral (or total energy) as
\begin{equation}
E_t = E_k + E_m + E_g + E_h.
\end{equation} 
When shocks appear, as in our simulations, this energy is not conserved
during the evolution (see also the simulations of Matsumoto \etal 1990).
The curve labeled $E_t$ in Figure~4 is constant up
to nearly the end of the linear stage, because the perturbation speeds are below $a$ at
those stages. When shocks begin to form during the nonlinear regime, the total energy
begins to decrease but it levels up later, when the gas speeds fall below the sound speed
and the system enters into the damping oscillatory stage. This is particularly clear in the
MS case. For the MA case, there are very weak shocks even at the damping oscillatory stage
and the total energy curve is still decreasing at $t=80$ (see lower right panel in Fig.~3).

The kinetic energy, $E_k$, in both MS and MA cases shows small variations,
even though shocks are generated during the nonlinear stage.  This is due to 
the fact that these shocks occur in low density regions.  
The heat-exchange term, $E_h$, corresponds to the amount of work done by
or on the gas (when $\Delta E_h$ is positive, work is done by the gas; see Mouschovias
1974). The gravitational energy, $E_g$, decreases as gas from high latitudes falls down to
the magnetic valleys. Comparing the runs for $E_g$ in both panels, one finds
that a larger amount of gravitational energy is released in the MA case. This is because
the MA oscillations are able to bend the midplane, and a larger amount of gas lowers its
position to be collected in the high density condensations (see Paper I). The
gravitational energy is first transformed into kinetic energy, and then it is either
dissipated in shocks or goes into magnetic field compression and tension. 
In general, the instability proceeds at the expense of lowering the
gravitational energy (Mouschovias 1974).

To compare the final equilibrium state derived in Paper I with the late times results of
the numerical simulations, we obtain the column densities as a function of $y$. For a given
horizontal location, the column density at any time $t$ is 
\begin{equation}
N(y, t) = \int_{z(y,A[z=0,t=0])}^{z_{\rm max}} \rho(y,z,t) dz,
\end{equation}
where the upper limit of the integral corresponds to the upper boundary of the
computational domain, and the lower one is the $z$-coordinate of the magnetic field line
that was initially located at midplane (labeled $A[z=0,t=0]$; see Paper I). During the
linear stage, the run of the column densities has only small variations. The important
changes occur during the nonlinear stage, when the contrast between condensations and voids
becomes maximum. Later on, during the oscillatory stage, the column densities maintain the
values achieved during the end of the nonlinear regime and the values at $t=80$, normalized
to the initial column density value $N(t=0)$, are plotted in Figure~5 for both MS (dotted
line) and MA (dashed line) perturbations. The MA case has the larger density contrast
between condensations and voids. As stated before, this is because the MA perturbations
provide more room for the high density gas by bending the midplane. At this last time, the
undular instability has almost reached the new final equilibrium in both MS and MA cases,
and the final column density configurations are very similar to those shown in Figure~5
of Paper I. The differences in the numerical and analytical results for the MS case are
of only about 10\%. The differences in the MA case are in general larger, particularly in
the voids, but they are also of the order of 10 \% at the center of the magnetic valleys.

In Paper I, we found that the growth rate increases, up to a certain maximum value, with
the location of the nodal point. This is because gravity, the driving force of the
instability, increases up to a certain maximum value. To verify this dependence in the
numerical experiments, we made a series of cases with different values for the
$z$-boundaries. Here we show one simulation, case 6, with an extended vertical domain, $-12
\le z \le 12$, and a sinusoidal MA perturbation. Figure~6 shows the resulting rms
velocities (dotted and dashed lines as in Fig. 2) as a function of time. The solid line
now has a slope equal to the growth rate corresponding to the nodal points for this case.
As is clear from the figure, the growth of the linear part follows the rate predicted by
the linear analysis for this case.   

\subsection{Random Perturbations} 

As a final issue of this part, we now verify the wavelength value and parity of the
fastest growing mode in the thick disk system. For this purpose we have performed
simulations with random velocity perturbations, without any preferred parity or
wavelength. The horizontal size of the computational domain has been extended, and the size
of the simulation in case 7 is three times longer than in the previous models. We also have
increased accordingly (by a factor of three) the number of cells in the $y$-axis.

Figure~7 shows the run of the rms velocities for this case and, as before, the slope of
the solid line corresponds to the growth rate. The horizontal component (dotted line)
has an initial adjustment and enters into the linear growth at $t\simeq 8$. The velocity
grows linearly, up to $t\simeq 45$, with the predicted maximum growth rate. As in the
previous models, the rms of the vertical component has some initial fluctuations and then, 
after $t\simeq 32$, it also grows with the predicted rate. The rest of the evolution is
similar to the one already described in case 5. The only difference is that the
evolutionary stages are now significantly delayed. The resulting structures are displayed
in two snapshots, at $t=45$ and $t=63$ (the end of the linear and nonlinear stages,
respectively), in Figure 8. The series of condensations are formed at slightly different
times, but they have the inter-distance scale of the most unstable wavelength and the MA
symmetry. This shows that, even when there was a parity degeneracy in the linear analysis,
the odd parity mode is dominant.

\section{DISCUSSION}

In this paper, we have confirmed numerically the results of the linear analysis obtained 
in Paper I for
the undular mode of the Parker instability in a thick, warm disk model of the Galaxy. 
The dependence of the growth 
rate on the position of the upper boundary of the numerical grid has been confirmed, and
an explanation in terms of the action of the gravitational field is provided. The main
properties of the final equilibrium configuration were also confirmed. The differences
between the final states for the symmetric and antisymmetric modes were corroborated by
the numerical study. The actual values of the column density distributions are slightly
different, but the shapes and the final-to-initial column density ratios were confirmed. 
In addition, the present study also provides a demonstration that the antisymmetric mode
is dominant and its signature should be present -- that of an odd parity with respect to
the central galactic plane -- in structures formed by the instability. 
In addition to the above, we are confident of the new evolutionary details of the 
instability in a multi-component thick disk, since we have got the same results
with two different numerical codes (TVD and ZEUS-3D). 


These new details may have important implications in our understanding of the galactic
disk. The extended vertical structure observed in our Galaxy and others, as the diffuse
ionized warm layer, forces higher scale heights (for mass, pressure and $B$-fields) than
assumed in previous calculations of the Parker instability. The resulting thick disk
alters considerably the time and length scales involved in the process and, as long as the
upper layers are not dominated by a hot coronal plasma phase, the disk is unstable to
undular modes. Evidence for the pervasiveness of this hot phase is still under debate 
(see Cox 1995, and the conference proceedings edited by Arthur, Brickhouse
\& Franco 2000) but, regardless of the actual details of the gas temperature distribution, one can
explore the consequences of the instability with the present isothermal thick disk model.

In considering some of the possible applications of this work, one has to
keep in mind the fact that our results refer only to a 2--D analysis. The
final equilibrium states, with a well defined periodicity of troughs and
crests, are unstable to 3--D perturbations (Asseo \etal 1978).
Nonetheless, tridimensional simulations in a thin disk (Kim \etal 1998)
show that some of the 2--D structuring is still present in this case.
The instability in this case is a combination of two modes, interchange
and undular. The interchange mode corresponds to the case in which the
field lines maintain their initial direction but some flux tubes are
displaced upwards when they find themselves less dense than the
surroundings. The undular mode (which is the one addressed in this study,
and is simply referred to as the Parker instability) corresponds to the
case in which the field lines are wavy above the midplane and gas
aggregation occurs at the magnetic valleys. Three dimensional motions are
more easily destabilized than those in two dimensions, for density
fluctuations (which instigate the interchange mode) are related to the
total (gas plus magnetic) pressure, whereas the undular mode depends on
the gas pressure alone (Hughes \& Cattaneo 1987).

The most unstable growth rate of the interchange mode has a vanishing
wavelength along the third dimension, which is perpendicular to both
gravity and initial field directions. This fact has been known from
one of Parker's (1967) pioneering works, and can limit the size and
mass of the resulting structures.
Our isothermal disk is unstable to the 3--D interchange mode (Martos
1993), and its fastest growth rate is that of an infinitesimally small
wavelength perpendicular to the 2--D plane of motion considered in our
calculations. As a result, and depending on the coupling of the field
lines, small scale structuring or turbulence will be also made. Additional
physical effects, such as rotation, self--gravity, or variable gravity in
the $z$ direction, cannot stabilize the buckling of the field lines
(Asseo \etal 1978; Elmegreen 1991). Thus, the instability will tend to
proceed faster in 3--D than in 2--D, creating complex small scale
structuring, but some of the 2--D signatures will be also present at the
large, kiloparsec size scales (\ie the large scale undular mode, with its
odd parity with respect to the central galactic plane, will persist). The
relative importance of the undular large scale structures depends on
whether one is considering arm or inter-arm regions. The inter-arm regions
are sheared, and the effects of rotation in the interchange mode lead to
a preferential small scale structuring (see Kim, Ryu \& Jones 2000). Thus,
little or no undular signatures are found, and alternating dense and
rarefied sheets are formed by the three-dimensional perturbations
(Kim \etal 1998, 2000). The density structure of these sheets are far
from those expected for giant molecular clouds, and the enhancement
factors for the vertical column density are small. Thus, it is
difficult to consider the Parker instability alone as the formation
mechanism of giant clouds in the general interstellar medium (the
resulting 3--D structures may be better more reminiscent of the worms
observed in atomic hydrogen; see Koo, Heiles \& Reach 1992).

In contrast, inside the arms, with a stronger gravitational field and
no shear, the large scale undular structures are clearly present
along the arms (Franco \etal 2000). The spacing between the resulting
condensations is similar to the most unstable mode derived here, and
the antisymettric mode creates a corrugation pattern along the spiral
arm. An interesting test of this prediction is suggested by the
structuring detected in the Carina arm, with a corrugation scale of
2.4 kpc (Alfaro, Cabrera-Ca\~no \& Delgado 1992). Early 3--D studies
already showed that the spiral density wave triggers the instability
in the thick gaseous disk (Martos \& Cox 1994), and current work in
3--D (Franco \etal 2000) indicates that the undular mode indeed gives
rise to a corrugation, or bending, along the arm. There is good
agreement with the observed corrugation length when appropriate
scaling of the gravitational field towards the inner Galaxy is
accounted for. This feature, along with the hydraulic jumps generated
by the action of spiral density waves (Martos \& Cox 1998), can
create strong spatial and velocity distortions in the vicinity of
spiral arms. These issues certainly require further observational
studies of corrugations and the velocity fields across spiral arms
(Alfaro \etal 2000).

Physics not included in this study will alter some details of the picture
discussed above. For instance, the Galactic field is mostly tangled and
the random component must have a stabilizing influence (Zweibel \& Kulsrud
1975). Also, self--gravity (Elmegreen 1982) and twisting of the field
lines (Shibata \& Matsumoto 1991; Kim, Ryu \& Jones 2000) will contribute
to the formation of coherent, finite (but small) size structures, setting
more realistic lower limits to the theoretical zero wavelength predicted
by theory for the fastest growing interchange mode. In addition, magnetic
reconnection should have played an important role at the late evolutionary
stages. The twisted random field components may lead to a rapid energy
release (\eg Lazarian \& Vishniac 1999), and can create additional flows
and modify the final energy budget of the process. We notice, however,
that the scales of the problem addressed here are much larger than the
cell size of the random component, 50 to 100 pc. The effect of galactic
differential rotation, as stated above, will modify the outcome of the
instability (\eg Foglizzo \& Tagger 1994, 1995). Finally, the timescales
for our 2--D simulations are much larger than the cosmic ray residence
time of 2.$\times 10^7$ years, making a causal association between cosmic
ray escape and the Parker instability unlikely. But again, this, as most
previous issues, demands precise three--dimensional studies, such as those
presented by Kim \etal (1998) and Kim, Ryu \& Jones (2000), and should be
addressed in the near future.

\acknowledgments
We warmly thank Gene Parker and Don Cox for useful comments and
suggestions during the development of this project, and Jane Arthur for a
careful reading of an early version of this paper. The comments and
suggestions made by the referee are gratefully acknowledged. We also thank
M. Norman, M. MacLow and R. Fielder for continued advice on ZEUS-3D. This
work has been partially supported by a bilateral CONACYT-Mexico and
KOSEF-Korea agreement. JF, MM and AS acknowledge partial support from
DGAPA--UNAM grant IN130698, and by a R\&D grant from Cray Research Inc.
The work by JK was supported in part by the Office of the Prime Minister
through Korea Astronomy Observatory grant 99-1-200-00. The work by SSH was
supported in part by grant BSRI-98-5411. The work by DR was supported in 
part by the grant KRF-99-015-DI0113. The numerical
calculations were performed using UNAM's ORIGIN-2000 supercomputer and SUN
Enterprise 3500 at the Korea Astronomy Observatory.  


\newpage

\begin{deluxetable}{ccccccc}
\tablecolumns{7}
\tablecaption{Parameters of the 2-D runs}
\tablewidth{0pt}

\tablehead{
\colhead{Case} & 
\colhead{Computational Domain} &
\colhead{Resolution} &
\colhead{Perturbation} &
\colhead{Parity} &
\colhead{$\lambda_y$} &
\colhead{$\lambda_z$} \\
\colhead{} & \colhead{[$H$]} & \colhead{[cells]} & \colhead{} &
\colhead{} & \colhead{[$H$]} & \colhead{[$H$]} 
}

\startdata

1 & $-9 \le z \le 9$ & 128 & no & - & - & - \nl

2 & $-9 \le z \le 9$ & 256 & no & - & - & - \nl

3 & $-9 \le z \le 9$ & 512 & no & - & - & - \nl

4 & $-9 \le y \le 9, ~~0 \le z \le 9$ & 256$\times$128 & sinusoidal 
& MS & 18 & 18 \nl

5 & $-9 \le y \le 9, -9 \le z \le 9$ & 256$\times$256 & sinusoidal 
& MA & 18 & 36 \nl

6 & $-9 \le y \le 9, -12 \le z \le 12$ & 192$\times$256 & sinusoidal 
& MA & 18 & 36 \nl


7 & $-27 \le y \le 27, -9 \le z \le 9$ & 768$\times$256 & random 
& - & - & - \nl

\enddata
\end{deluxetable}

\clearpage
 
		FIGURE CAPTIONS

Fig. 1.- The panels show the maximum velocity of case 1 as a function
of time. The upper plot shows the results obtained with the ZEUS code,
and the lower plot shows the same case performed with the TVD code. The
velocities and times are in units of $a$ and $H_{\rm eff}/a$,
respectively.

Fig. 2.- The frames show the run of the rms velocities for the undular
instability. The panels display the rms value of the horizontal (dotted
lines) and vertical (dashed lines) velocities, as a function of time,
for both MS (left panels) and MA (right panels) perturbations. The
upper frames were obtained with the TVD code, and the lower ones with
ZEUS-3D. The solid lines have a slope equal 
to the growth rate for the fastest growing mode with $z_{\rm node}= 9$ 
derived in Paper I.

Fig. 3.- The Parker instability in 2-D for a magnetized
multi--component gaseous disk, performed with the TVD code. The
sequence shows the density (color
logarithmic scale), velocity field (arrows) and magnetic fields
(lines), at three selected times: $t$ = 35, 45, 80 for the MS mode
(case4), and $t$ = 27, 36, 80 for the MA mode (case5). 

Fig. 4.- The time evolution of different energies (internal $E_i$;
kinetic $E_k$; magnetic $E_m$; gravitational $E_g$; heat-exchange
$E_h$, and total $E_t$) for the simulations presented in Fig. 3. All
energies are normalized to their initial values. The left panel
corresponds to the MS case, and the right side to the MA case.

Fig. 5.- Final column density values as a function of horizontal
location, for the MS (dotted line) and MA (dashed line) cases of Fig.~3. 
The column densities are normalized to the initial value, 
$N(t=0)$.

Fig. 6.- Same as in Fig. 2, but for case 6 performed with the TVD code. The
slope of the solid line represents the growth rate corresponding to the
fastest growing mode with nodal point $z_{\rm node}= 12$ obtained in
Paper I.

Fig. 7.- Same as in Fig. 2, but for case 7 performed with the TVD code. The
slope of the solid line is the same as in Fig. 2.

Fig. 8.- Parker instability in a magnetized multi--component gaseous
disk for case 7. The sequence shows the density (color logarithmic
scale), velocity field (arrows) and magnetic fields (lines), at two
selected times: $t$ = 45, 63.


\clearpage
 
\begin{thebibliography}{}

\bibitem[]{}Alfaro, E. J., Cabrera-Ca\~no, J., \& Delgado, A. J. 1992,
\apj, 399. 576

\bibitem[]{}Alfaro, E. J., P\'erez, E., Gonz\'alez Delgado, R. M., Franco,
J. \& Martos, M. A. 2000, in preparation

\bibitem[]{}Arthur, S. J, Brickhouse, N., \& Franco, J. (ed.) 2000,
Astrophysical Plasmas: Codes, Models and Observations, RevMexAA
(Conf. Ser.), in press

\bibitem[]{}Asseo, E.,\& Sol, H. 1978, PhysRep 48, No. 6, 307, 206

\bibitem[]{}Baierlein, R. 1983, \mnras, 205, 669

\bibitem[Basu et al. 1997]{BMP97}
Basu, S., Mouschovias, T. Ch., \& Paleologou, E. V. 1997, \apj, 480, L55


\bibitem[Bienaym\'e, Robin \& Crez\'e 1987]{BRC}Bienaym\'e, O., Robin, A.C., \& 
Crez\'e, M. 1987, A\&A, 180, 94

\bibitem[Boulares \& Cox 1990]{BC}
Boulares A., \& Cox, D. P. 1990, ApJ, 365, 544 (BC)

\bibitem[]{}
Cox, D. P. 1995, Nature, 375, 185

\bibitem[Cox 1990]{C90}Cox, D. P. 1990, in Interstellar Disk--Halo Connection
in Galaxies, ed. H. Bloemer (Kluwer), 143

\bibitem[Dettmar 1990]{}Dettmar, R.--J. 1990, A\&A, 232, L15

\bibitem[]{}Elmegreen, B.G. 1982, \apj, 253, 634


\bibitem[]{}
Evans, C. R., \& Hawley, J. F. 1988, \apj, 332, 659

\bibitem[]{}
Foglizzo, T., \& Tagger, M. 1994, \aap, 287, 297 

\bibitem[]{}
Foglizzo, T., \& Tagger, M. 1995, \aap, 301, 293 

\bibitem[]{}Franco, J., Santill\'an, A., \& Martos, M. A. 1995, in
Formation of the Milky Way, ed. E. Alfaro \& A. Delgado, Cambridge Univ. Press,
97

\bibitem[]{}Franco, J., Kim, J., Alfaro, E. J. \& Hong, S. S. 2000, in
preparation

\bibitem[]{}
Heiles, C. 1996, ApJ, 462, 316

\bibitem[]{}Horiuchi, T., Matsumoto, R., Hanawa, T., \& Shibata, K. 1988, 
PASJ, 40, 147

\bibitem[Hoyle \& Ellis 1963]{HE}Hoyle, F., \& Ellis, G.R.A. 1963,  Australian 
JPhys, 16, 1       

\bibitem[Howk \& Savage 1997]{HS97}
Howk, J.C., \& Savage, B. 1997, AJ, 114, 2463

\bibitem[]{}Hummel, E., \& Beck, R. 1995, A\&A, 303, 691

\bibitem[Kaisig \etal 1990]{kai90}
Kaisig, M., Tajima, T., Shibata, K., Nozawa, S., \& Matsumoto, R.
1990, \apj, 358, 698

\bibitem[Kalberla \& Kerp 1998]{KK98}
Kalberla, P. M. W., \& Kerp, J. 1998, \aap, 339, 745

\bibitem[]{} Kim, J., Franco, J., Hong, S.S., Santill\'an, A., \&
Martos, M.A. 2000, \apj, 531, 873

\bibitem[]{} Kim, J., Hong, S. S., Ryu, D. \& Jones, T. W., 1998, \apj,
506, L139

\bibitem[]{} Kim, J., Ryu, D., \& Jones, T. W. 2000, in Astrophysical
Plasmas: Codes, Models and Observations, RevMexAA Conf. Ser., in press

\bibitem[]{} Kim, J., Ryu, D., Jones, T. W., \& Hong, S. S. 1999, \apj,
514, 506

\bibitem[]{} Koo, B.-C., Heiles, C. \& Reach, W. T. 1992, \apj, 390, 108

\bibitem[]{} Lazarian, A., \& Vishniac, E. T. 1999, ApJ, 517, 700

\bibitem[Martos 1993]{M93}Martos, M. A. 1993, Ph.D. Thesis, UW-Madison

\bibitem[]{}
Martos, M. A., \& Cox, D. P. 1994, Numerical Astrophysics, ed. J. Franco, S. Lizano, L. Aguilar \& E. Daltabuit, Cambridge Univ. Press,
229

\bibitem[]{}
Martos, M. A., \& Cox, D. P. 1998, ApJ, 509, 703

\bibitem[]{}
Martos, M. A., Allen, C., Franco, J., \& Kurtz, S. E. 1999, ApJ, 526, L89

\bibitem[Matsumoto et al. 1990]{MHSH88}
Matsumoto, R., Horiuchi, T., Shibata, K, \& Hanawa, T. 1988, \pasj, 40, 171

\bibitem[Matsumoto et al. 1990]{MHHS90}
Matsumoto, R., Horiuchi, T., Hanawa, T., \& Shibata, K. 1990, \apj, 356, 259

\bibitem[Matsumoto \& Shibata 1992]{MS92}
Matsumoto, R., \& Shibata, K. 1992, \pasj, 44, 167

\bibitem[]{}Mouschovias, T. Ch. 1974, \apj, 192, 37

\bibitem[]{}Mouschovias, T. Ch., Shu, F. H., \& Woodward P. R. 1974, A\&A, 
33, 73

\bibitem[Nozawa \etal 1992]{noz91}
Nozawa, S., Shibata, K., Matsumoto,  R., Sterling, A. C.,  Tajima, T.,
Uchida, Y.,  Ferrari, A., \& Rosner, R. 1992, \apjs, 78, 267

\bibitem[Parker 1966]{P66} Parker, E.N. 1966, ApJ, 145, 811

\bibitem[Rand 1995]{} Rand, R. 1995, AAS, 187, 4811

\bibitem[Rand, Kulkarni \& Hester 1989]{RKH} Rand, R., Kulkarni, S., \& Hester,
 J. 1990, ApJ, 362, L35

\bibitem[Reynolds 1989]{RR}Reynolds, R.J. 1989, ApJ, 339, L29

\bibitem[]{}Ryu, D., Miniati, F., Jones, T. W., \& Frank, A. 1998, \apj, 
509, 244

\bibitem[]{}Shibata, K., \& Matsumoto, R. 1991, Nature, 353, 633

\bibitem[]{}Santill\'an, A., Franco, J., Martos, M.A., \& Kim, J., 1999, 
\apj, 515, 657

\bibitem[Shibata \etal 1989a]{shi89a}
Shibata, K., Tajima, T., Matsumoto, R., Horiuchi, T., Hanawa, T.,
Rosner, R., \& Uchida, Y. 1989a, \apj, 338, 471

\bibitem[Shibata \etal 1989b]{shi89b}
Shibata, K., Tajima, T., Steinolfson, R. S., \& Matsumoto, R.
1989b, \apj, 345, 584.

\bibitem[Stone \& Norman 1992a]{SNa}Stone, J.M., \& Norman, M.L. 1992a, ApJS, 
80, 753

\bibitem[Stone \& Norman 1992b]{SNb}Stone, J.M., \& Norman, M.L. 1992b, ApJS, 
80, 791


\bibitem[Valle\'e 1997]{V97}Valle\'e, J.P., 1997, in Fundamental of Cosmic Physics, 19, 1 

\bibitem[]{}Zweibel, E. G., \& Kulsrud, R. M. 1975, \apj, 201, 63

\end{thebibliography}
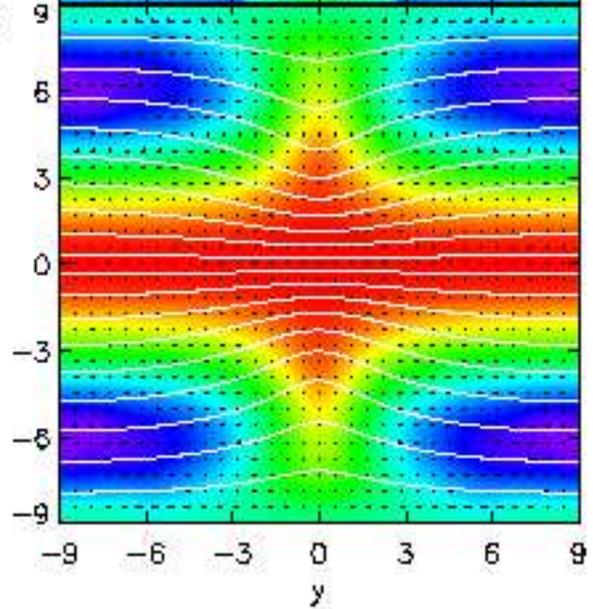
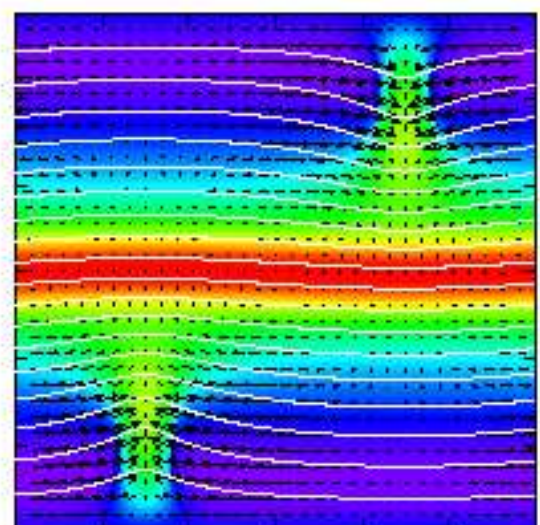
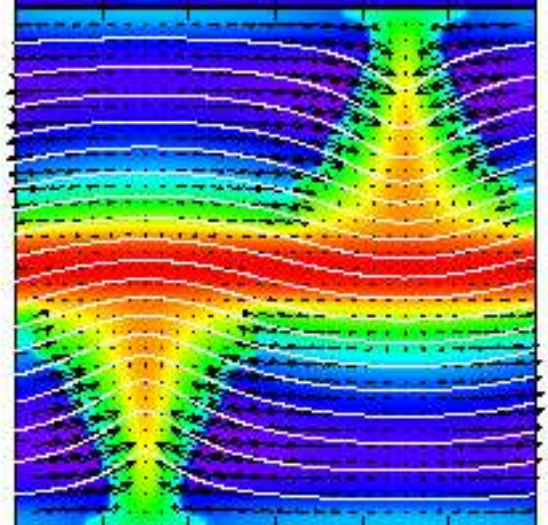
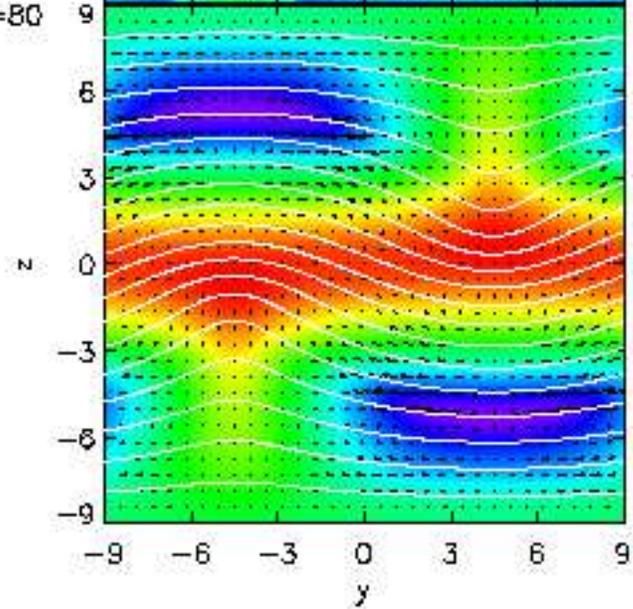
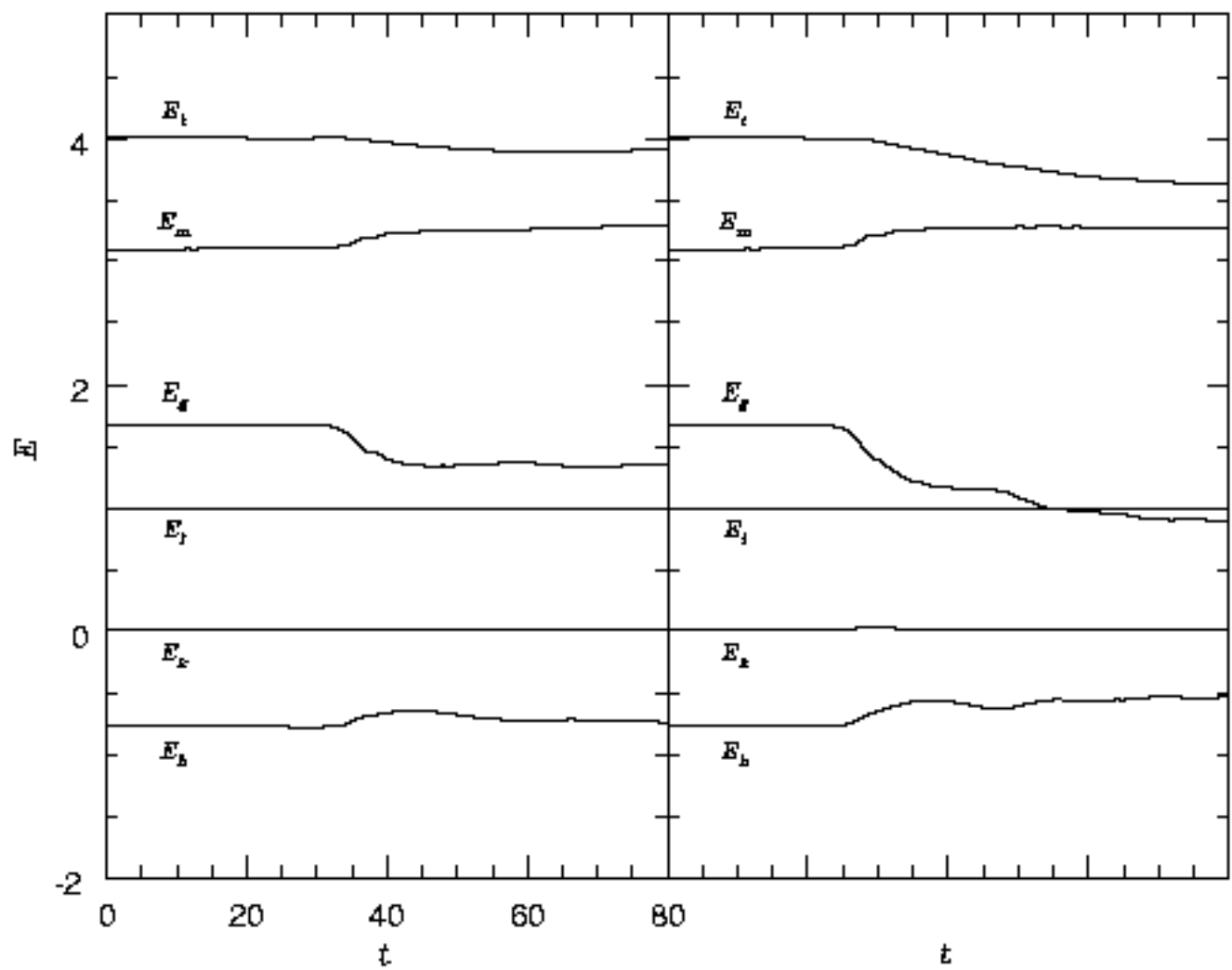
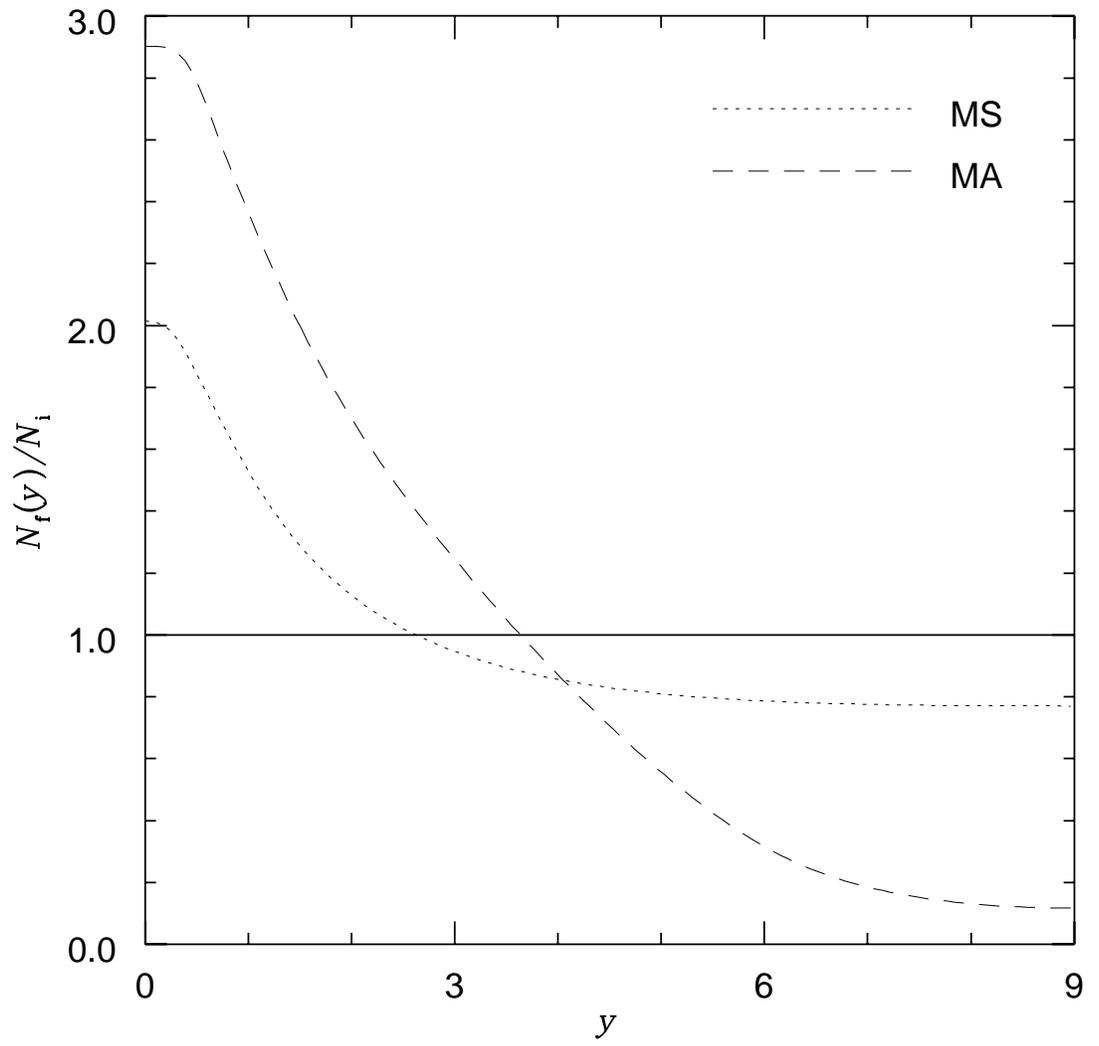
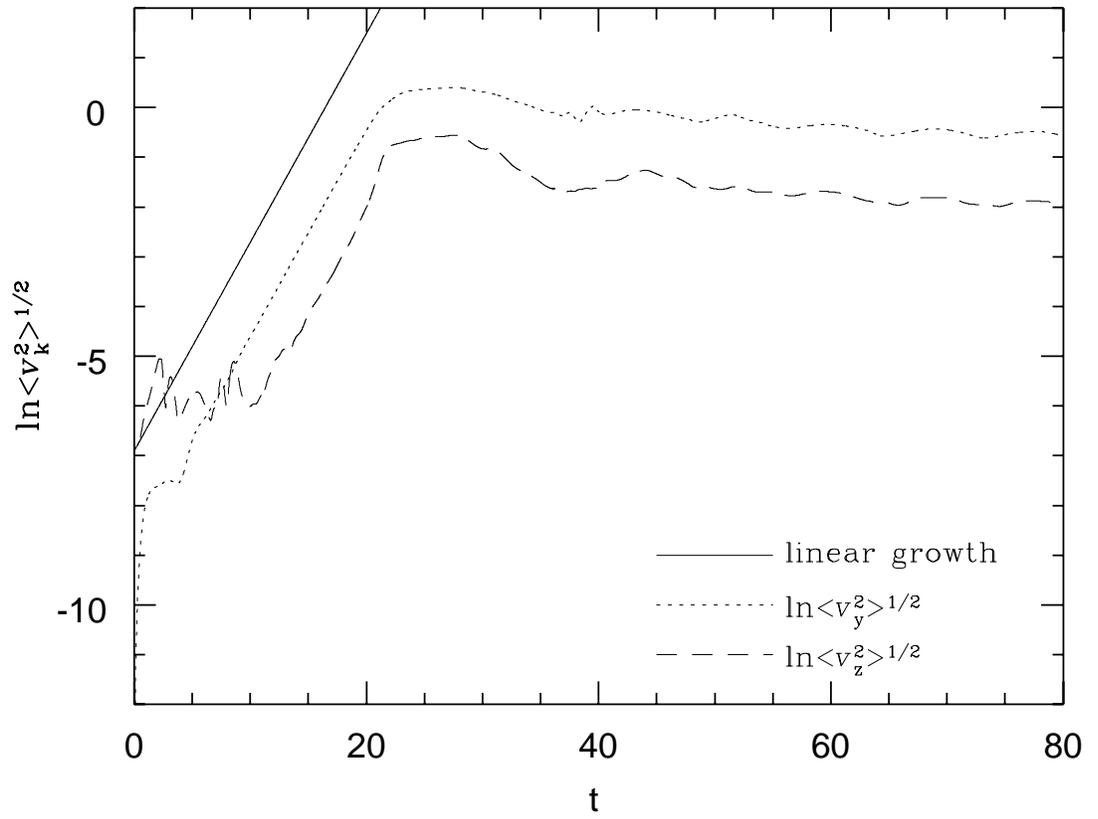
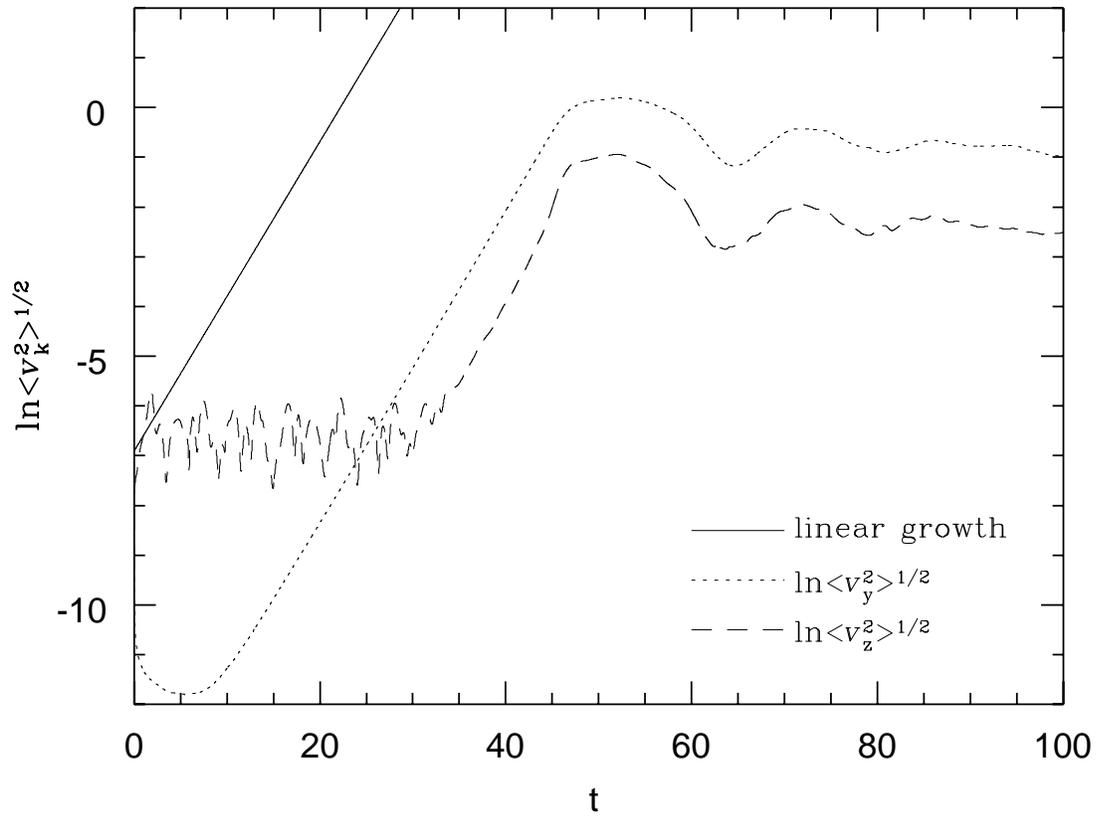
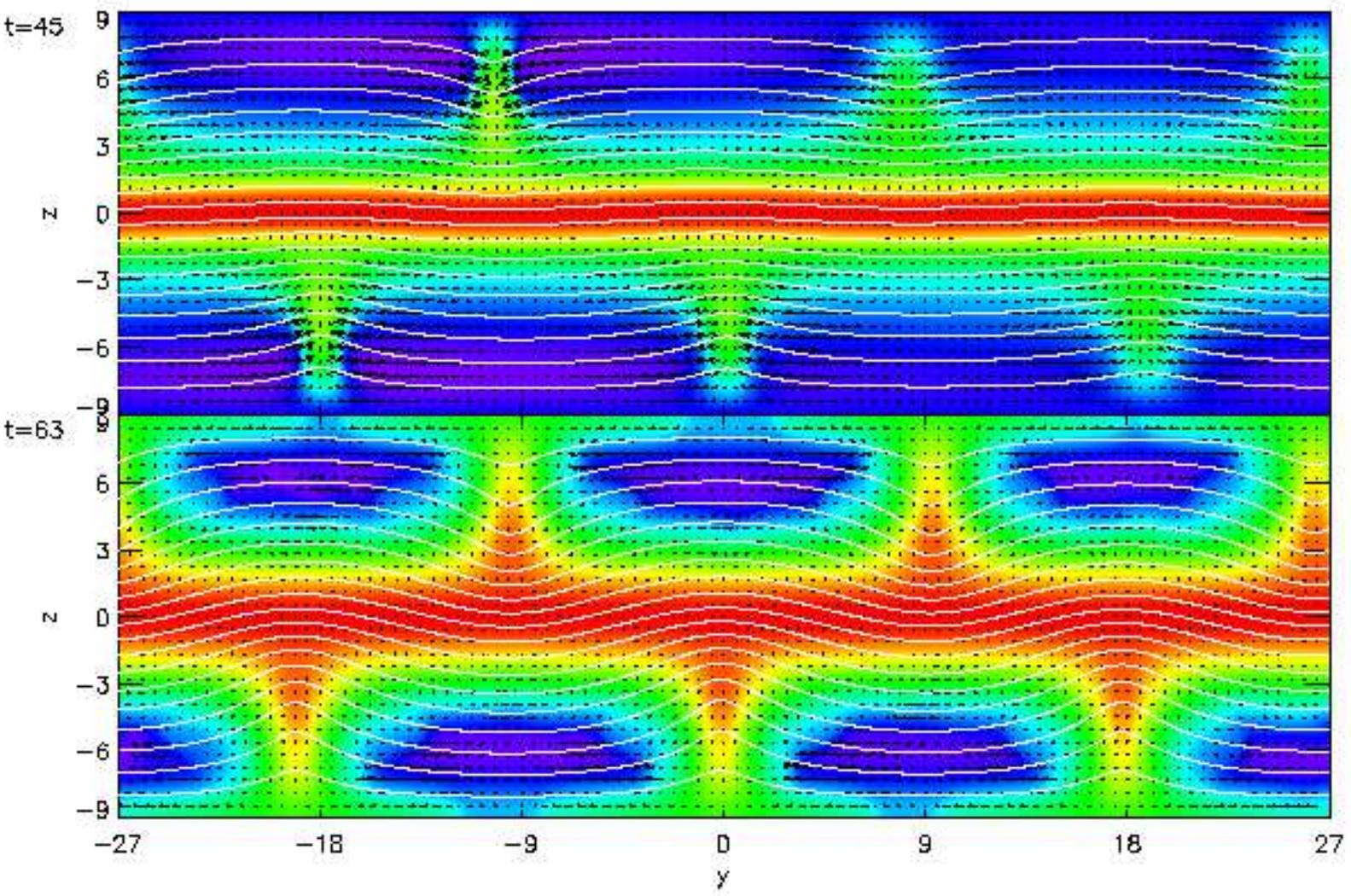
\end{document}